\documentclass[%
 reprint,
 amsmath,amssymb,
 aps,
]{revtex4-2}

\usepackage{graphicx}% Include figure files
\usepackage{subfigure} 
\usepackage{dcolumn}% Align table columns on decimal point
\usepackage{bm}% bold math
\usepackage{float}
\usepackage{color}
\usepackage{ulem}
\usepackage{amsmath}
\begin{document}

\preprint{APS/123-QED}

\title{Critical behavior of Non-Hermitian Kondo effect in a pseudogap system}% Force line breaks with \\

\author{Jiasong Chen}
\affiliation{%
 National Laboratory of Solid State Microstructures and Department of Physics \\
 Nanjing University, Nanjing 210093, China
}%

\date{\today}% It is always \today, today,

\begin{abstract}
A combined study of non-Hermitian physics and strong correlations can yield numerous intriguing effects. Authors of a previous study on the non-Hermitian Kondo model in the ultracold atoms reported the reversion of the renormalization group (RG) flow. In this paper, we investigate the non-Hermitian Kondo effect in the system with a specific form of density of states $\rho (\omega) \sim |\omega|^{r}(r>0)$, known as the pseudogap system. We find that, for $r<\frac{1}{2}$, our results from perturbative RG are consistent with those of the Hermitian pseudogap Kondo effect. For $r=\frac{1}{2}$, a fixed point with the reversion property emerges in the RG flow. For $r>\frac{1}{2}$, an unstable fixed point appears in the complex plane of the parameter space. Furthermore, through the large-$N$ expansion, we validate the RG results for $r < \frac{1}{2}$, finding that the self-consistent equations have non-trivial solutions in a specific region of the complex plane of the parameter space.
% \begin{description}
% \item[keywords]
% Kondo effect, Non-Hermitian physics, Perturbative renormalization group, Large N expansion
% \item[PACS]
% 72.15.Qm;72.10.Fk;71.55.-i;
% \end{description}
\end{abstract}
% \keywords{Suggested keywords}%Use showkeys class option if keyword

\maketitle

%\tableofcontents

\section{Introduction}

The Hermitian nature of the Hamiltonian is one of the key assumptions in quantum mechanics. It ensures the conservation of probability in isolated quantum systems and guarantees that the energy expectation value of a quantum system is real. However, in practical quantum systems, there are often exchanges of energy, particles, and information with the environment. Consequently, open quantum systems can often give non-Hermitian terms in Hamiltonians  \cite{daley2014quantum, luschen2017signatures, rotter2009non}.  Researchers have shown that non-Hermitian physics can lead to many effects, such as the non-Hermitian skin effect \cite{yao2018edge,zhang2022review,okuma2020topological,borgnia2020non}, non-Hermitian proximity effect \cite{wu2023effective}, spontaneous parity-time symmetry breaking resulting in imaginary parts in particle spectra \cite{bender1998real,heiss2012physics}, unusual phase transitions \cite{guo2022emergent,li2023non,yu2024non}, and some topological states \cite{yao2018non,bergholtz2021exceptional,kawabata2019symmetry}. Subsequently, researchers also investigated some non-Hermitian strongly correlated electron models \cite{nakagawa2018non,lourencco2018kondo,
han2023complex,kulkarni2022kondo,yoshida2018non,yamamoto2019theory}. In cold atom experiments, the long-lived metastable excited states of alkaline earth atoms can act as local magnetic moments, while their ground states play the role of conduction electrons \cite{riegger2018localized,gorshkov2010two,kanasz2018exploring,kuzmenko2018multipolar}. In Ref. \cite{nakagawa2018non}, the authors proposed that the two-body losses/gains induced by inelastic scattering between the ground and excited states give rise to the non-Hermitian Kondo effect. Using the perturbative renormalization group (RG) , they found that, due to the introduction of non-Hermitian terms, the RG flow exhibits a reversion at the weak-coupling fixed point. Specifically, the RG flow of the system originates from the zero value of the parameters and subsequently returns to it.

At low temperature, magnetic impurities strongly affect the properties of conduction electrons. The Kondo effect \cite{hewson1997kondo,kondo1964resistance} is characterized by a temperature scale $T_{K}$: When the temperature ($T$) is larger than 
 $T_{K}$, the conduction electrons are only weakly coupled with the impurity; for $T < T_{K}$ the (antiferromagnetic) coupling grows nonperturbatively and leads to the formation of a many-body singlet with the conduction electrons, which completely screens the impurity magnetic moment. 

Furthermore, Dirac and Weyl semimetals have attracted widespread attention due to their unique electronic properties. In these materials, the conduction and valence bands touch at isolated points, named Dirac points and Weyl points, in the Brillouin zone. Near these isolated points, the conduction electrons exhibit a pseudogap density of states (DOS) characterized by $\rho(\omega) \sim |\omega|^{r} (r>0)$. Additionally, in $d$-wave superconductors, due to the form of the energy gap function, the superconducting gap vanishes at specific nodes, where excitations with Dirac conelike dispersion emerge. The unique band structures of these materials often give rise to numerous properties. Moreover, their interactions with localized magnetic moments have attracted significant interest \cite{fritz2013physics,withoff1990phase,principi2015kondo,khadka2020kondo,mitchell2015kondo,fritz2004phase,kircan2004critical,gonzalez1998renormalization,cassanello1996kondo,da2009tunable,ma2018kondo,balatsky2006impurity,polkovnikov2001impurity}. In Ref. \cite{withoff1990phase}, the author initially investigated the coupling effects between the conduction electrons with the pseudogap DOS and the localized magnetic moment based on poor man's scaling and mean-field theory. The results revealed a critical parameter controlling the phase transition between decoupled and strong-coupling phases at zero temperature exists. Subsequently, in Ref. \cite{gonzalez1998renormalization}, the authors studied the phase diagram of the pseudogap Kondo problem by numerical RG (NRG) theory. They discovered that, in systems with particle-hole symmetry, an unstable fixed point, denoted as $J_{c}$, emerges on the RG flow when $r < \frac{1}{2}$. This fixed point, termed the symmetric critical (SCR) fixed point, separates the local-moment (LM) phase from the symmetric strong-coupling (SSC) phase. When the Kondo coupling coefficient $J < J_{c}$, the RG flow moves toward the LM fixed point, while for $J > J_{c}$, it flows toward the SSC fixed point. This conclusion is consistent with the results obtained from mean-field theory. When $r > \frac{1}{2}$, the SCR fixed point merges with the SSC fixed point. In this case, the system does not exhibit a strong-coupling phase, regardless of the strength of the coupling coefficient. Subsequently, detailed studies have been conducted on the Kondo model for $r <0$ \cite{fritz2013physics,mitchell2013quantum,vojta2002fractional,mitchell2013kondo,zhuravlev2011kondo,shankar2023kondo}. These models, characterized by a divergent DOS, exhibit rich phase transitions in both ferromagnetic and antiferromagnetic couplings \cite{vojta2002fractional,mitchell2013quantum}. Such models may appear in systems like graphene with vacancies \cite{mitchell2013kondo} and materials with van Hove singularities \cite{zhuravlev2011kondo}. The DOS of magic-angle graphene often exhibits van Hove singularities when the angle between the two graphene layers is close to the magic angle. In recent years, as interest in the magic-angle graphene system has grown, these models have also been extensively studied \cite{shankar2023kondo,chang2024vacancy}.

In this paper, we investigate the critical behavior of the non-Hermitian Kondo effect in a pseudogap system to study the response of the pseudogap materials to localized magnetic moments in the open quantum system. For $r < \frac{1}{2}$, through perturbative RG, we find that in the complex plane of the parameter space, the system partially flows to the LM fixed point and partially to the SSC fixed point. This behavior suggests there is a quantum phase transition between the LM phase and the SSC phase in complex plane.  For $r=\frac{1}{2}$, the LM and SSC fixed points merge and generate a fixed point with the reversion property. When  $r>\frac{1}{2}$, the SSC fixed point disappears and a new unstable fixed point emerges in the complex plane. Subsequently, we employ the large-$N$ expansion method and define the residual function of the self-consistent equations. We find that the nontrivial solutions of the self-consistent equations exist in a specific region of the complex plane, thereby confirming the conclusion of the perturbative RG for $r<\frac{1}{2}$. 

The paper is organized as follows. In Sec. \ref{se model}, we introduce the non-Hermitian Kondo model in a pseudogap system. In Sec. \ref{se rg}, we obtain the RG flow through perturbative RG. As a result, we identify four distinct patterns of the RG flow for different values of $r$. In Sec. \ref{se N}, by the large-$N$ expansion, we obtain and solve the self-consistent equations of the system. By defining the residual function of the equations, we identify the interval in which a nontrivial solution exists. In Sec. \ref{se co}, we summarize the results of this paper and discuss their implications.

\section{Model}
\label{se model}
In this paper, we start with ultracold Fermi atomic gas with a pseudogap DOS in a three-dimensional optical lattice. The expression for this DOS is as follows:
\begin{equation}
    \rho (\omega)  = 
\begin{cases}
 C|\omega |^{r} & \text{ if } |\omega| \le  D \\
 0 & \text{ else } ,
\end{cases}
\end{equation}where $\omega$ and $D$ respectively represent the energy and bandwidth of the conduction fermions and $C$ is the prefactor. The expression for the Hamiltonian of the kinetic energy of conduction fermions is as follows \cite{fritz2004phase}:
\begin{equation}
    H_{c}= \sum _{\sigma } \int dk |k|^{r} k c^{\dagger }_{k \sigma} c_{k \sigma} ,
\end{equation}
where $c_{k \sigma }$ is the annihilation operator of the conduction fermion with momentum $k$ and spin $\sigma $. 

In the experiment of ultracold fermionic atoms, some fermions in this cold atomic gas absorb energy and transition to an excited state, becoming localized in a small spatial region through the specific wavelength, behaving like a local impurity, while the fermionic atoms in the ground state act as conduction electrons \cite{riegger2018localized,gorshkov2010two,foss2010probing,nakagawa2018non}. The Hamiltonian of this entire cold atomic gas system is described by the Anderson model as follows:
\begin{equation}
\begin{aligned}
    H_{A} =  H_{c} + t_{A} \sum_{k \sigma} ( c^{\dagger }_{k \sigma} f_{\sigma} + \mathrm{H.c.} ) \\  
    + E_{f} \sum_{\sigma} n_{f \sigma} + U_{r} n_{f\uparrow} n_{f\downarrow},
    \label{An}
\end{aligned}
\end{equation}where $f_{\sigma}$ is the annihilation operator of local fermions with spin $\sigma $, and $n_{f\sigma } = f^{\dagger}_{\sigma}f_{\sigma}$ is the number operator of local fermions. In Eq. (\ref{An}), the second term describes the transition between the conduction and local fermions with amplitude $t_{A}$. The third term describes the energy levels of the local fermions, and the fourth term represents the interaction with strength $U_{r}$ between them.

So far, we have formulated the Hermitian pseudogap Anderson model. By the Lindblad master equation, we account for the coupling between the system and the environment, which introduces a non-Hermitian term into the Hamiltonian . Finally, through the Schrieffer-Wolff transformation, we derive the expression for the Hamiltonian of the non-Hermitian pseudogap Kondo effect (see Appendix \ref{ap}):
\begin{equation}
    H_{K} = H_{c} + J \vec{S}_{\mathrm{imp}} \cdot \vec{s}(0).
    \label{eq.nk}
\end{equation}
Here, $\vec{S}_{\mathrm{imp}}$ is the impurity spin, and $\vec{s}(0)$ is the spin of conduction fermions at the point of local fermions. The Kondo coupling parameter is a complex number, i.e., $J = J_{r} + iJ_{i}$.

\section{renormalization group}
\label{se rg}
\begin{figure*}
\subfigure[]{
\includegraphics[scale=0.34]{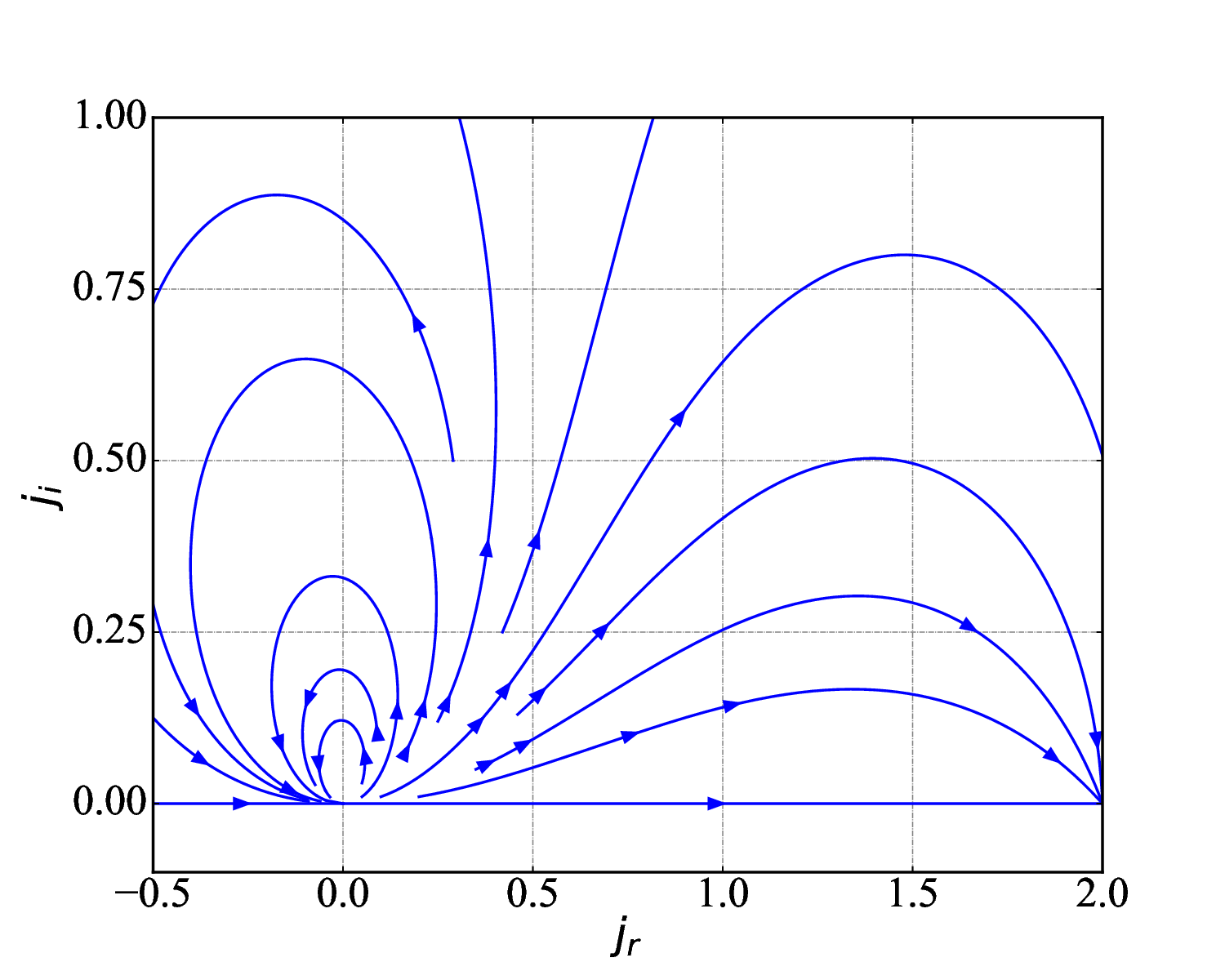} \label{0}
}
\subfigure[]{
\includegraphics[scale=0.34]{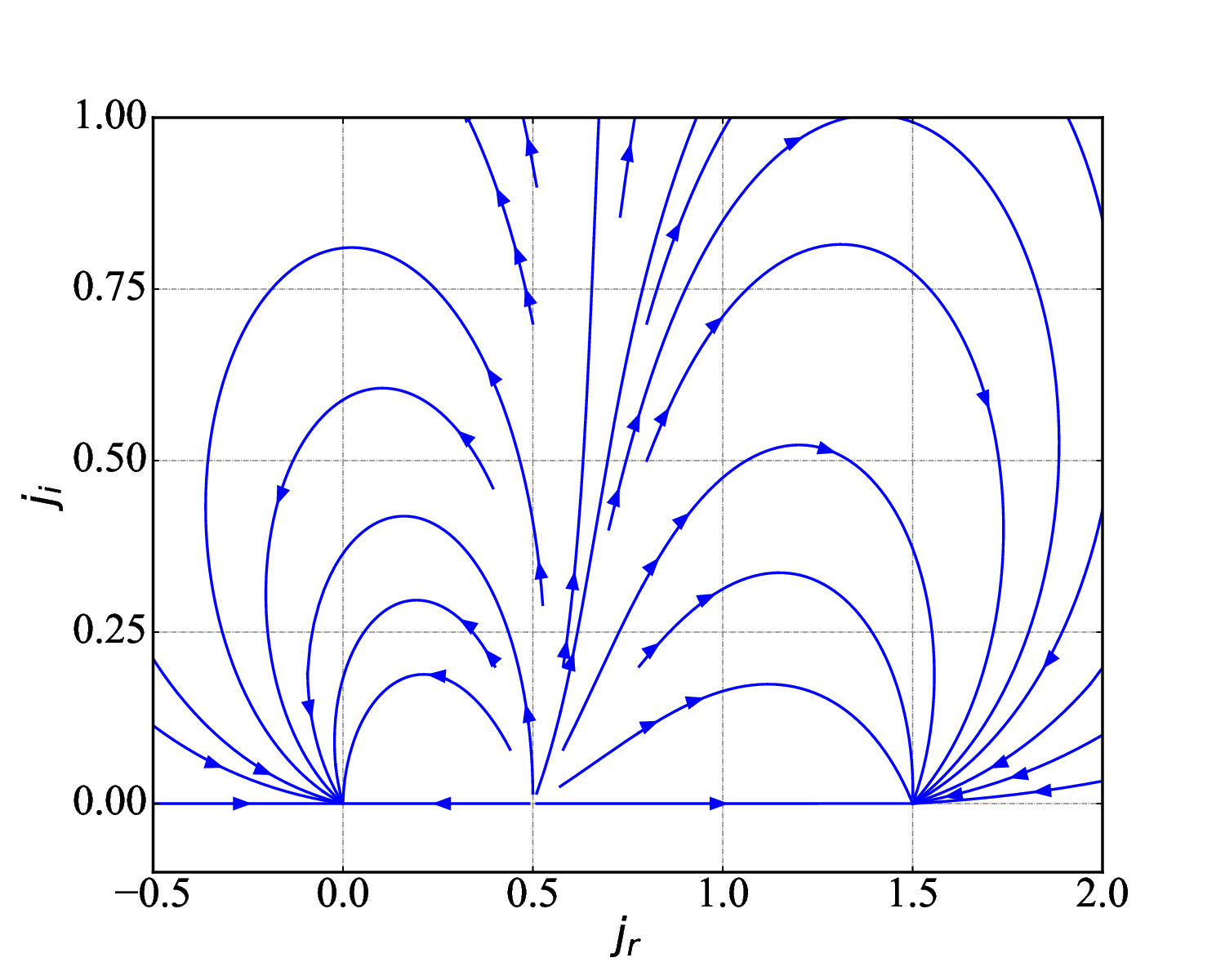} \label{1}
}\\
\subfigure[]{
\includegraphics[scale=0.34]{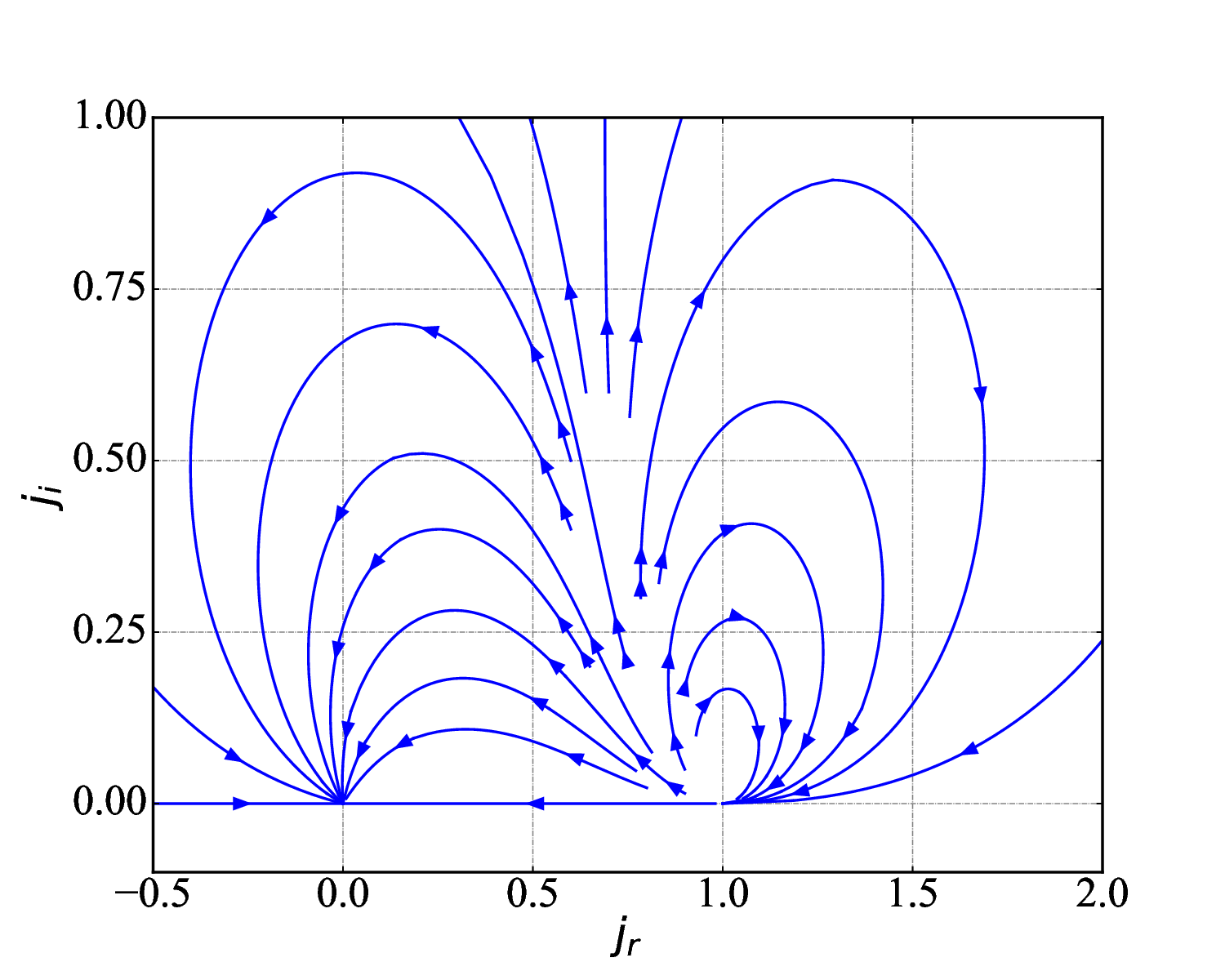} \label{2}
}
\subfigure[]{
\includegraphics[scale=0.34]{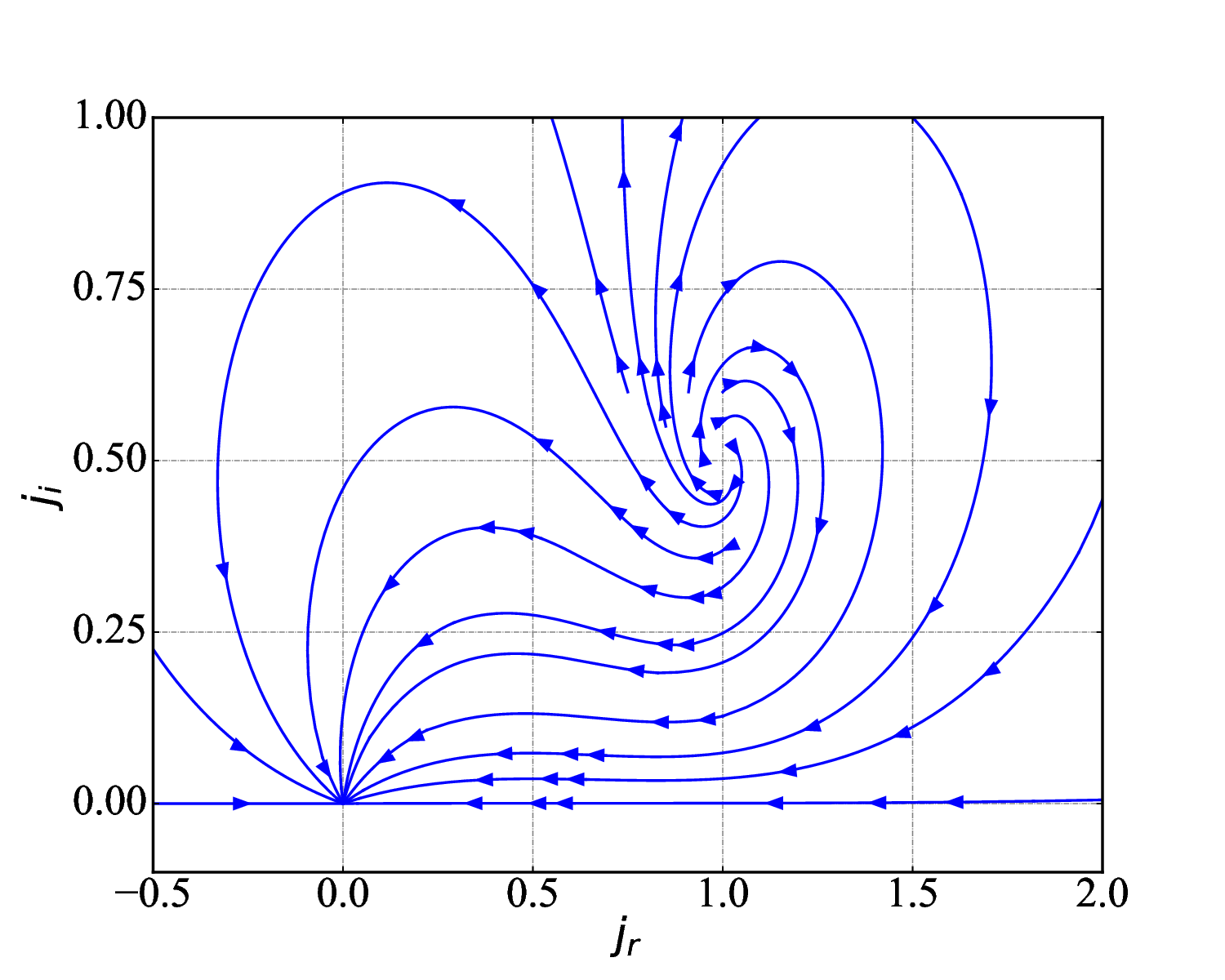} \label{3}
}
\caption{Renormalization group (RG) flow diagrams of the non-Hermitian pseudogap Kondo model 
in different values of $r$: (a) $r=0$, (b) $r=0.375$, (c) $r=0.5$, and (d) $r=0.625$.}
\label{RG flow}
\end{figure*}

We calculate the perturbative RG equation for the Hamiltonian in Eq. (\ref{eq.nk}) up to the third order and derive the following form:
\begin{equation}
   \frac{ \mathrm{d} j}{\mathrm{d}\ln{D}} = rj - j^{2} + \frac{j^{3}}{2} ,
\end{equation}
where $j = \rho(D) J$ is the dimensionless coupling constant. This expression can be regarded as a direct extension of the RG equation for the Hermitian pseudogap Kondo effect \cite{fritz2013physics, nakagawa2018non}. By analysis, we find that there are some different patterns of the RG flow, as shown in Fig. \ref{RG flow}.

First, we consider the special case of $r=0$, where the DOS is constant and independent of the energy of the conduction fermions. At this point, we recover the non-Hermitian metallic Kondo effect, which has been discussed in detail in Ref. \cite{nakagawa2018non}. As shown in Fig. \ref{0}, in the complex plane, part of the RG flow moves toward the strong-coupling fixed point [located at (2.0,0)], while another part flows to the LM fixed point, which exhibits the reversion property.

For $0<r<\frac{1}{2}$, as shown in Fig. \ref{1}, with the increase of $r$, the LM fixed point, which originally exhibited the reversion property, now splits into two fixed points: LM and SCR, while the latter is an unstable fixed point [located at (0.5, 0)] that controls the phase transition between the LM and the SSC phases [located at (1.5,0)] on the real axis. In the complex plane, part of the RG flow moves toward the SSC fixed point, while another part flows toward the LM fixed point, suggesting that two different types of phases exist in the complex plane. The specific form of the critical point $j_{c}$ on the real axis is as follows:
\begin{equation}
   j_{c} = 1 - \sqrt{1-2r} \approx  r + O(r^2) .
\label{jc}
\end{equation}

For $r = \frac{1}{2}$, as shown in  Fig. \ref{2}, the critical points of the SCR and SSC phases merge. This leads to the emergence of a fixed point with the reversion property, as mentioned in Ref. \cite{nakagawa2018non}, at the SSC fixed point. In other words, the RG flow originates from the point $(1.0, 0)$ and returns to the same point.

For $r>\frac{1}{2}$, the RG flow is shown in Fig. \ref{3}. For the Hermitian case ($J_{i}=0$) within this interval, there is no SSC fixed point on the real axis, and therefore, no phase transition occurs between the LM and SSC phases. In the complex plane, a new unstable fixed point emerges. The RG flow of the system moves from this point toward the LM fixed point. This new fixed point suggests that the introduction of non-hermiticity may cause the system to produce some intriguing critical behaviors in the complex plane.

\section{Large-$N$ Expansion}
\label{se N}

To prepare for the large-$N$ expansion, we first extend the Kondo model Hamiltonian to one with SU$(N)$ symmetry, referred to as the Coqblin-Schrieffer (CS) model \cite{coleman2015introduction}:

\begin{equation}
    H_{\mathrm{CS}} = H_{c} 
    - \frac{J}{N}\sum_{k , \alpha  }\sum_{k^{\prime} , \beta  } c^{\dagger}_{k  \alpha }f_{\alpha}f^{\dagger}_{\beta}c_{k^{\prime}  \beta  } ,
\label{CS}
\end{equation}
where $\alpha, \beta = 1, 2, \dots , N$ is the extended spin quantum number of fermions and $J$ is the complex Kondo coupling parameter.

Subsequently, in the path integral formulation, it is required to introduce a complex Lagrange multiplier $\lambda=\lambda_{r}+i\lambda_{i}$ \cite{yamamoto2024correlation} to impose the single occupancy condition for the local fermions $n_{f}= \sum_{\alpha}f^{\dagger}_{\alpha}f_{\alpha} = 1$ .

The next step is to apply the Hubbard-Stratonovich transformation to rewrite the four-operator interaction term in Eq. (\ref{CS}) into the following form:

\begin{equation}
    \begin{aligned}
                - \frac{J}{N} &\sum_{k , \alpha  }\sum_{k^{\prime} , \beta  } c^{\dagger}_{k\alpha }f_{\alpha}f^{\dagger}_{\beta}c_{k^{\prime}  \beta  }
       \\  &\rightarrow 
     \sum_{k , \alpha } (\bar{V}c^{\dagger}_{k  \alpha}f_{\alpha} + V f^{\dagger}_{\alpha} c_{k \alpha})  
        + N\frac{\bar{V}V}{J} ,
    \end{aligned}
\end{equation}
where 
\begin{equation}
\begin{aligned}
        V &= \sum _{\alpha }{} _{\mathrm{L}}  \langle c^{\dagger}_{k\alpha}f_{\alpha}  \rangle _{\mathrm{R}} \\       
          &=   \sum _{\alpha ,n}{} \frac{ _{\mathrm{L}}  \langle E_{n} | c^{\dagger}_{k\alpha}f_{\alpha} | E_{n} \rangle _{\mathrm{R}} \exp{(-\beta E_{n})} }{Z},
\end{aligned}
\end{equation}
\begin{equation}
\begin{aligned}
        \bar{V} &= \sum _{\alpha } {} _{\mathrm{L}} \langle f^{\dagger}_{\alpha}c_{k\alpha} \rangle _{\mathrm{R}} \\
    &=   \sum _{\alpha ,n}{} \frac{ _{\mathrm{L}}  \langle E_{n} |f^{\dagger}_{\alpha}c_{k\alpha} | E_{n}\rangle _{\mathrm{R}} \exp{(-\beta E_{n})}}{Z}
\end{aligned}
\end{equation}
are the static boson fields describing the coupling between the local fermions and conduction fermions. Here, $\beta$ is a parameter used to formulate the path integral. Also, $\left | E_{n} \right \rangle_{\mathrm{L}}$ and $\left | E_{n} \right \rangle_{\mathrm{R}}$ are the eigenvectors of $H_{CS}$ with eigenvalue $E_{n}$, satisfying $H^{\dagger}_{\mathrm{CS}} \left | E_{n} \right \rangle_{\mathrm{L}} = E^{*}_{n} \left | E{n} \right \rangle_{\mathrm{L}}$ and $H_{\mathrm{CS}} \left | E_{n} \right \rangle_{\mathrm{R}} = E_{n} \left | E_{n} \right \rangle_{\mathrm{R}}$, respectively. Further, $Z=\sum_{n} \exp{(-\beta E_n)}$ is the partition function. Since $H_{\mathrm{CS}}$ is non-Hermitian, we have $\left | E_{n} \right \rangle_{\mathrm{L}} \ne \left | E_{n} \right \rangle_{\mathrm{R}}$, and thus $\bar{V} \ne V^{*}$. However, due to the U$(1)$ gauge symmetry of our system, we can choose a specific gauge such that $V = \bar{V} = V_{r} + iV_{i}$ \cite{yamamoto2019theory}.

Finally, in the limit $N \rightarrow \infty$, the result of the path integral is entirely determined by the saddle point of the integrand. In summary, we arrive at the following expression for the free energy $F$ in large-$N$ limit:

\begin{equation}
\begin{aligned}
    F = &-\frac{N}{\pi}\int ^{D}_{-D} \frac {d\omega}{\exp{(\beta \omega) }+1} 
\tan ^{-1}\left [ \frac{\pi C |\omega |^{r} V^{2}}{-\omega +\lambda - V^{2} C \mathcal{P}(\omega) 
}  \right ]
\\&+N\frac{V^{2}}{J} - \lambda  ,
\end{aligned}
\end{equation}

where $\mathcal{P}(\omega) = \mathcal{P} \int \frac{d\epsilon   |\epsilon |^{r}}{\omega -\epsilon } = \pi  |\omega|^{r}\tan(\frac{\pi r}{2})\mathrm{sgn} (\omega) $ represents the principal value  integral. In the saddle-point approximation, the parameter of free energy should satisfy the following self-consistent relations:

\begin{equation}
    \begin{aligned}
        \frac{\partial F}{\partial V} & = 0 , \\
        \frac{\partial F}{\partial \lambda } & = 0 .
    \end{aligned}
\end{equation}
At the limit of $\beta \rightarrow \infty$, the specific forms of these two equations are as follows:

\begin{equation}
\begin{aligned}
    \int^{0}_{-D} d\omega \frac{V C |\omega |^{r}(-\omega +\lambda)}{\left [-\omega +\lambda - V^{2} C \mathcal{P}(\omega ) \right ]^{2}+\pi V^{2} C |\omega |^{r} } 
    -\frac{V}{J} = 0 ,
\end{aligned}
\label{eq1}
\end{equation}
\begin{equation}
    \int^{0}_{-D} d\omega \frac{V^{2} C |\omega |^{r}}
{\left [ -\omega +\lambda - V^{2}C\mathcal{P}(\omega ) \right]   ^{2}+\pi V^{2} C |\omega |^{r} }
-\frac{1}{N} = 0 .
\label{eq2}
\end{equation}

By examining the self-consistent Eqs. (\ref{eq1}) and (\ref{eq2}), we can immediately deduce that $V = 0$ is a solution in the large-$N$ limit. Since all interactions are mediated by the bosonic field $V$, if this trivial solution is stable, it indicates the decoupling of the local fermions from the conduction fermions, corresponding to the LM phase. If nontrivial solutions to the equations exist, the system is in the SSC phase.

To discuss the nontrivial solutions of the equations, we first eliminate $V$ from Eq. (\ref{eq1}). Assuming the existence of a nontrivial solution with small $V^{2}$ and $\lambda$ near the critical point, we set  $V^{2}$, $\lambda$ to be equal to 0, after eliminating. This yields the expression for the critical Kondo coupling parameter $J_{c}$ on the real axis:

\begin{figure}

    \centering
    \includegraphics[scale = 0.3]{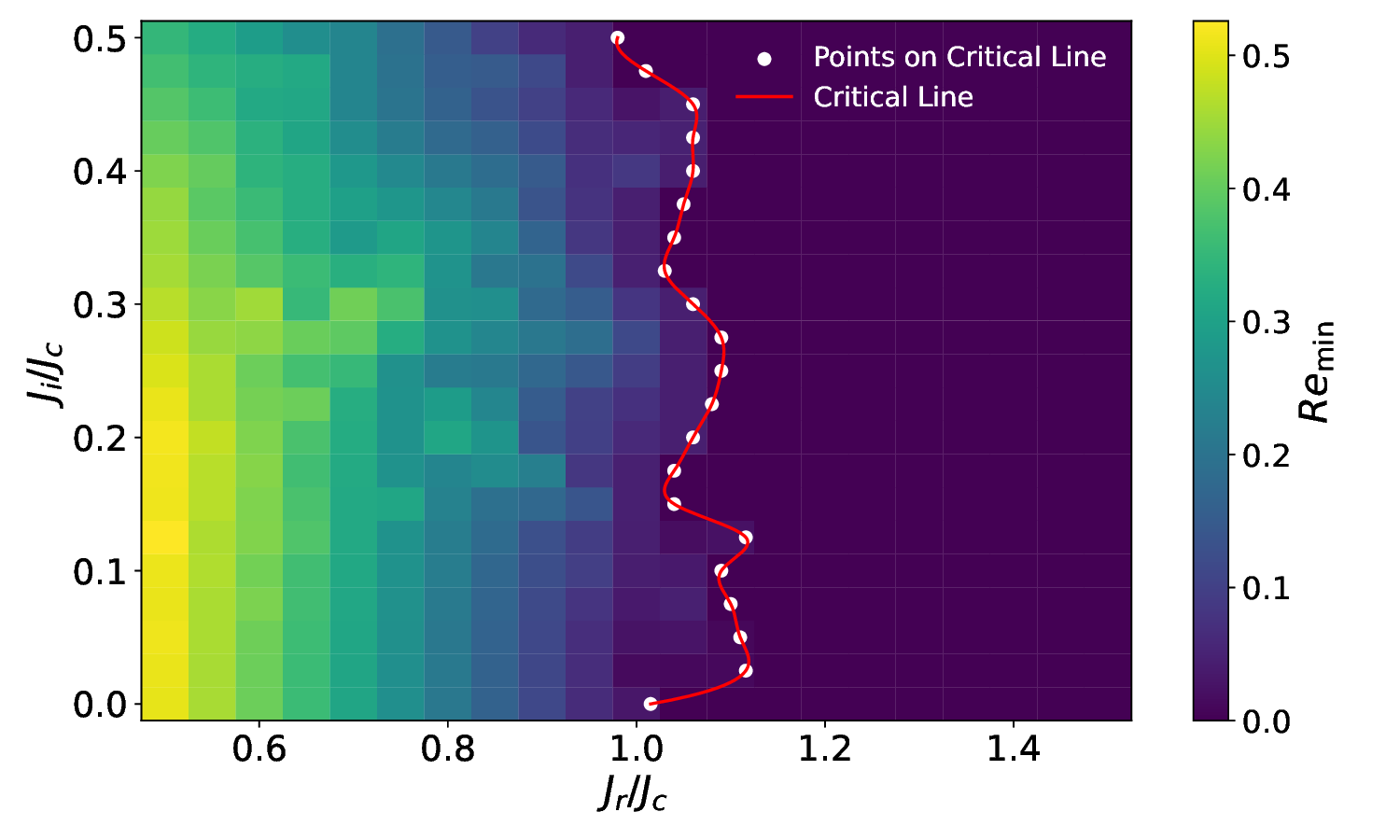} 
    \caption{ The distribution of function $\mathrm{Re}_{\text{min}}$ for $r = 0.4$, $D = 10^{4}$, and $N = 10$. It is evident that a critical line in $\mathrm{Re}_{\text{min}}$ starts from the critical point on the real axis, dividing the parameter space into two regions. The white points represent the coordinates on the critical line, and the red line is the critical line obtained through fitting. 
 }
\label{fig:surface}
\end{figure}

\begin{equation}
    J_{c} = \frac{r}{CD^{r}} .
\end{equation}
This result is consistent with Eq. (\ref{jc}) obtained from the RG theory for $r < \frac{1}{2}$.

% To derive the solutions of Eqs.(\ref{eq1}) and (\ref{eq2}), we define the residual function as follows:

Then, based on Eqs. (\ref{eq1}) and (\ref{eq2}), we define the residual function as follows:
\begin{equation}
    \mathrm{Re}(J_{r}, J_{i}, V_{r}, V_{i}, \lambda _{r}, \lambda_{i}) = \sqrt{ |f_{1} - 1|^{2} + |f_{2} - 1|^2 } ,
\end{equation}
where
\begin{equation}
        f_{1} =  JC \int^{0}_{-D} d\omega \frac{|\omega |^{r}(-\omega +\lambda)}{\left [ -\omega +\lambda - V^{2}\mathcal{P}(\omega ) \right ] ^{2}+\pi |\omega |^{r}V^{2} }  ,
\label{eq f1}
\end{equation}
\begin{equation}
        f_{2} = N \int^{0}_{-D} d\omega \frac{V^{2}|\omega |^{r}}
{\left [ -\omega +\lambda - V^{2}\mathcal{P}(\omega ) \right ] ^{2}+\pi |\omega |^{r}V^{2} } .
\label{eq f2}
\end{equation}
Here, we have replaced $V^{2}C$ with $V^{2}$, effectively absorbing the prefactor $C$ into the order parameter.

By adjusting the variables $V_{r}$, $V_{i}$, $\lambda _{r}$ and  $\lambda_{i}$, we can find the minimum of $Re$ for each Kondo coupling parameter, termed $\mathrm{Re}_\text{min}(J_{r}, J_{i})$. Here, we take $r=0.4$ as an example to demonstrate the properties of the function $\mathrm{Re}_{\text{min}}(J_{r}, J_{i})$ in the complex plane. The parameters used in the calculation are $D=10^{4}$ and $N=10$ (we define that $D$ is in unit of $J_{r}$, and $J_{r}$ is set to 1).  Fig. \ref{fig:surface} illustrates the surface of the function $\mathrm{Re}_{\text{min}}(J_{r}, J_{i})$ as it varies in the complex plane of the parameter space. Notice that in our calculations, the variables $J_{r}/J_{c}$ and $J_{i}/J_{c}$ are two dimensionless coupling coefficients. If we set $J_{r}/J_{c}=x$ and $J_{i}/J_{c}=y$, it follows that $J=J_{r}+iJ_{i}=J_{c}(x+iy)=\frac{r}{CD^{r}}(x+iy)$. When this expression is substituted into Eq. (\ref{eq f1}), the prefactor $C$ is canceled out. Therefore, the prefactor $C$ no longer appears in the numerical calculations.

For points where the function $\mathrm{Re}_{\text{min}} = 0$, the non-trivial solutions to our equations exist; conversely, for points where $\mathrm{Re}_{\text{min}} \neq 0$, the non-trivial solutions do not exist. 

Through our calculations, we find that within the interval $0 < r < 1$, the function $\mathrm{Re}_{\text{min}}(J_{r},J_{i})$ has zeros in the complex plane, as shown in Fig. \ref{fig:surface}. A critical line (red line) emerges from the critical point on the real axis, dividing the complex plane into two distinct regions. On the left side of this line, where $\mathrm{Re}_{\text{min}}(J_{r}, J_{i}) \ne 0$, the self-consistent equations only have trivial solutions, placing the system in the LM phase. On the right side of the line, $\mathrm{Re}_{\text{min}}(J_{r}, J_{i})=0$ , signifying that the self-consistent equations have non-trivial solutions and the system transitions into the SSC phase. This result is consistent with the conclusions from the RG flow in Fig. \ref{1}. However, the result of mean-field theory is unreliable for $r > \frac{1}{2}$, in comparison with Ref. \cite{gonzalez1998renormalization}.

\section{conclusion}
\label{se co}

In summary, we derived the RG flow of the non-Hermitian Kondo effect in the pseudogap system through perturbative RG theory. The result reveals that when $r<\frac{1}{2}$, there exists an unstable fixed point (SCR) on the real axis, controlling the phase transition between the LM and SSC phases; when $r=\frac{1}{2}$, the SCR fixed point merges with the SSC phase, forming a fixed point with the reversion property; when $r>\frac{1}{2}$, an unstable fixed point appears in the complex plane. Subsequently, we derived the nontrivial solutions of the self-consistent equations in the complex plane, validating the conclusion of the perturbative RG for $r<\frac{1}{2}$. However, for $r > \frac{1}{2}$, the physical significance of this unstable fixed point in the complex plane remains elusive. To address this question, some exotic approaches such as non-Hermitian conformal field theory may be required.

Although the perturbative RG flow suggests that the system approaches a strong coupling regime, it is important to note that the perturbative theory becomes invalid before reaching this region. Thus, one cannot definitively claim that this represents a stable fixed point; the conclusion is merely suggestive. For instance, the two-channel Kondo model exhibits a perturbative RG flow at weak coupling similar to the one-channel model, yet the strong coupling fixed point is unstable in the former. This highlights the inherent limitations of the perturbative approach in fully characterizing the strong coupling physics. To fully understand the nature of the strong coupling regime, nonperturbative approaches would be highly valuable. Techniques such as NRG or other nonperturbative methods could be required to offer deeper insights. 

\begin{acknowledgments}
We are grateful to Rui Wang for helpful discussions. This work was supported by the National Natural Science Foundation of China (No. 12322402, No. 12274206).
\end{acknowledgments}


\appendix

\section{Derivation of Non-Hermitian Kondo Hamiltonian in the Pseudogap System}
\label{ap}

In this study, the cold atomic gas trapped by the three-dimensional optical lattice with a pseudogap DOS is the system under investigation, while the untrapped high-energy modes of the cold atoms, coupled to this system via inelastic collisions, serve as the environment. The system undergoing the two-body losses due to the inelastic collisions is described by the Lindblad master equation \cite{daley2014quantum, durr2009lieb}
\begin{equation}
    \frac{\mathrm{d} \rho }{\mathrm{d}t}= -i \left [  H, \rho  \right ] +   L_{f}\rho L^{\dagger}_{f} - \frac{1}{2} \left \{   L^{\dagger}_{f} L_{f} ,\rho   \right \},  
    \label{lq1}
\end{equation} where $\rho$ is the density matrix of the trapped atomic gas, and $L_{f}$ is the Lindblad operator describing the the two-body losses due to the interaction with the environment.
 The expression of $L_{f}$ is as follows \cite{yamamoto2019theory}: 
\begin{equation}
    L_{f} = \sqrt{\gamma} f_{\uparrow} f_{\downarrow} .
    \label{lf}
\end{equation}In the above equation, $\gamma > 0$ is defined as the loss rate.

By defining the effective non-Hermitian Hamiltonian $H_{\mathrm{eff}} = H_{A} - \frac{i}{2}  L^{\dagger}_{f} L_{f}$, we can rewrite Eq. (\ref{lq1}) as
\begin{equation}
    \frac{\mathrm{d} \rho }{\mathrm{d}t}= -i \left (  H_{\mathrm{eff}} \rho - \rho H^{\dagger}_{\mathrm{eff}}  \right ) +  L_{f} \rho L^{\dagger}_{f} ,
\end{equation}where the first term on the right-hand side represents the Schrödinger evolution under the effective non-Hermitian Hamiltonian and the second term describes the stochastic quantum jump process. The last term leads to the disappearance of the impurity, so the dynamics of the impurity is entirely described by the first term. By substituting Eq. (\ref{lf}) into $H_{\mathrm{eff}}$, we obtain that
\begin{equation}
    H_{\mathrm{eff}} =  H_{c} + t_{A} \sum_{k \sigma} ( c^{\dagger }_{k \sigma} f_{\sigma} + \mathrm{H.c.} ) + E_{f} \sum_{\sigma} n_{f \sigma} + U n_{f\uparrow} n_{f\downarrow},
\end{equation}which is a non-Hermitian Anderson Hamiltonian in the pseudogap system with complex interaction $U = U_{r} - i \gamma /2$. 

By Schriffer-Wolf transformation, we obtain the expression of the non-Hermitian Kondo Hamiltonian in the pseudogap system
\begin{equation}
    H_{K} =H_{c} + J\vec{S}_{\mathrm{imp}} \cdot \vec{s}(0) +  W \sum_{\sigma , \sigma ^{\prime }} c^{\dagger}_{\sigma}(0) c_{\sigma}(0) f^{\dagger}_{\sigma ^{\prime}} f_{\sigma ^{\prime}} ,
\label{nhH}
\end{equation}
where 
\begin{equation}
    J = t_{A}^{2} \left ( \frac{1}{E_{f} + U} - \frac{1}{E_f}  \right ) ,
\end{equation}\begin{equation}
    W = -\frac{1}{2} t_{A} ^{2} \left (  \frac{1}{E_{f} + U} + \frac{1}{E_f} \right) .
\end{equation}

In Eq. (\ref{nhH}), the second term on the right-hand side is the complex Kondo interaction, and the third term is the potential scattering term, where $c_{\sigma}(0) = \int dk \left | k \right |^{r} c_{k \sigma}$, $\vec{s}(0)=\frac{1}{2} \sum_{\sigma , \sigma^{\prime}}  c^{\dagger}_{\sigma ^{\prime}}(0) \vec{\sigma}_{\sigma ^{\prime} \sigma}c_{\sigma}(0) $, and $\vec{S}_{\mathrm{imp}} = \frac{1}{2} \sum_{\sigma , \sigma^{\prime}}  f^{\dagger}_{\sigma ^{\prime}} \vec{\sigma}_{\sigma ^{\prime} \sigma} f_{\sigma} $ ($\vec{\sigma}$ is the Pauli vector). Here, we only consider the case where the system has particle-hole symmetry, neglecting the effect of the potential scattering term.

Let $J = J_{r} + iJ_{i}$, and the explicit expressions for $J_{r}$ and $J_{i}$ are as follows:

\begin{equation}
    J_{r} = t_{A}^{2} \left [  \frac{ E_{f} + U_{r} }{(E_{f} + U_{r})^2 + \gamma ^{2}/4} - \frac{1}{E_{f}} \right ] ,
\end{equation}\begin{equation}
    J_{i} = t_{A}^{2} \left [ \frac{\gamma }{2(E_{f} + U_{r})^{2} + \gamma ^{2} /2} \right ] .
\end{equation} 
From the above equation, we can see that $J_{i}$ is always $>0$. Therefore, in the main text, we only consider the case when $J_{i}>0$.

\nocite{*}

\bibliography{apssamp}% Produces the bibliography via BibTeX.

\end{document}